\documentclass[twocolumn,showpacs,preprintnumbers,amsmath,amssymb]{revtex4}
\usepackage{graphicx}
\usepackage{dcolumn}
\usepackage{bm}
\begin{document}

\title{Scaling of geometric phases close to quantum phase transition in the XY chain}
\author{Shi-Liang Zhu} \affiliation{FOCUS Center and MCTP, Department
of Physics, University of Michigan, Ann Arbor, MI 48109.}

\begin{abstract}
We show that geometric phase of the ground state in the XY model
obeys scaling behavior in the vicinity of a quantum phase
transition. In particular we find that geometric phase is
non-analytical and its derivative with respect to the field strength
diverges at the critical magnetic field. Furthermore, universality
in the critical properties of the geometric phase in a family of
models is verified. In addition, since quantum phase transition
occurs at a level crossing or avoided level crossing and these level
structures can be captured by Berry curvature, the established
relation between geometric phase and quantum phase transitions  is
not a specific property of the XY model, but a very general result
of many-body systems.
\end{abstract}


\maketitle

\newpage

The phase factor of a wave function is the source of all
interference phenomena and one of most fundamental concepts in
quantum physics. Recent considerable interest in this field is
motivated by the pioneer work of Berry\cite{Berry}. Berry discovered
that a geometric phase, in addition to the usual dynamic phase, is
accumulated on the wave function of a quantum system, provided that
the Hamiltonian is cyclic and adiabatic. Since then, the adiabatic
geometric phase and its generalizations \cite{Aharonov,Sjoqvist}
have found many applications to broad fields \cite{Shapere,Bohm},
such as condensed matter physics \cite{Thouless,Niu,Morpurgo},
atomic, molecular, and optical physics, and quantum
computation\cite{Zanardi}, etc..

Very recently, Carollo and Pachos demonstrated the close relation
between geometric phases and quantum criticality of spin
chains\cite{Carollo}. In particular they showed that a nontrivial
geometric phase difference between the ground state and the first
excited state exists in the XX model if and only if the closed
evolution path circulates a region of criticality.  Quantum phase
transitions usually occur for a parameter region where the energy
levels of the ground state and the excited state cross or have an
avoided crossing, and is certainly one of the major interests in
condensed matter physics\cite{Sachdev,Wen,Barber} and quantum
information\cite{Osterloh}. Geometric phase, as a measure of the
curvature of Hilbert space, can reflect the energy level structures
and can capture certain features of quantum phase transitions.
However, at least two important problems need to be addressed. (i)
The XY model is parameterized by $\gamma$ and $\lambda$ (see the
definitions below Eq.(\ref{Hamiltonian})). Two distinct critical
regions appear in parameter space: the segment
$(\gamma,\lambda)=(0,(0,1])$ for the XX chain and the critical line
$\lambda_c=1$ for the whole family of the XY
model\cite{Sachdev,Lieb}. The nontrivial geometric phase difference
between the ground state and the first excited states calculated in
Ref. \cite{Carollo} can be used as a measure of the presence of the
first critical region, but this measure is unlikely to remain valid
for the second critical region.
Whether one can reveal the latter
critical region using geometric phase is of significance. (ii) As
also noted in Ref.\cite{Carollo}, a challenging but also important
question warranting further study is whether the typical features of
quantum criticality, such as the scaling feature, critical exponents
and universality, etc., have relation to geometric phases in this
many body system. Answering these questions is certainly significant
for a deeper understanding of quantum phase transitions, and also
from the perspective of geometric phases. So further results that
bridge these two interesting areas of research are of great
relevance.

In this paper, instead of using the geometric phase difference
between the ground state and the first excited state as a signature
of quantum criticality, we focus on the relation between geometric
phase of the ground state and quantum criticality in the XY chain.
We analyse geometric phases near the critical point of the XY model
and find that the geometric phase is non-analytical and its
derivative with respect to the field strength $\lambda$ diverges at
the critical line described by $\lambda_c=1$. In particular the
geometric phase obeys scaling behavior in the vicinity of a quantum
phase transition. Furthermore, universality in the critical
properties of geometric phase for a family of models is verified.
These results show that the key ingredients of quantum criticality
are present in geometric phases of the ground state. In addition, we
show that the relation between geometric phase and quantum phase
transitions established here is not model dependent, but is valid in
a wide variety of systems.

The system under consideration is a spin-1/2 XY chain, which
consists of N spins with nearest neighbor interactions and an
external magnetic field. The Hamiltonian of the system is given by
\begin{equation}
\label{Hamiltonian}  H=-\sum_{j=-M}^M \left
(\frac{1+\gamma}{2}\sigma_j^x\sigma_{j+1}^x+\frac{1-\gamma}{2}\sigma_j^y\sigma_{j+1}^y+\lambda\sigma_j^z
\right), \end{equation} where  $M=(N-1)/2$ for $N$ odd and
$\sigma^\mu_j\ (\mu=x,y,x)$ are the Pauli matrices for the $j$th
spin. We assume periodic boundary conditions. The parameter
$\lambda$ is the intensity of the magnetic field applied in the $z$
direction, and $\gamma$  measures the anisotropy in the in-plane
interaction. This XY model encompasses two other well-known spin
models: it turns into transverse Ising chain for $\gamma=1$ and the
XX (isotropic XY) chain in a transverse field for $\gamma=0$.

As for quantum criticality in the XY model, we need to distinguish
two universality classes depending on the anisotropy $\gamma$. The
critical features are characterized in term of a critical exponent
$\nu$ defined by $\xi \sim |\lambda-\lambda_c|^{-\nu}$ with $\xi$
representing the correlation length. For any value of the anisotropy
$\gamma $, quantum criticality occurs at a critical magnetic field
$\lambda_c=1$. For the interval $0<\gamma\le 1$ the models belong to
the Ising universality class characterized by the critical exponent
$\nu=1$, while for $\gamma=0$ the model belongs to the XX
universality class with $\nu=1/2$ \cite{Lieb,Sachdev}.

To investigate the geometric phase in this system, we introduce a
new family of Hamiltonians that can be described by applying a
rotation of $\phi$ around the $z$ direction to each spin, {\sl
i.e.}, $H_\phi=g_\phi H g_\phi^\dagger$ with $g_\phi=\prod_{j=-M}^M
\exp(i\phi\sigma_l^z/2)$\cite{Carollo}. The critical behavior is
independent of $\phi$ as the spectrum $\Lambda_k$ (see below) of the
system is $\phi$ independent. This class of models can be
diagonalized by means of the Jordan-Wigner transformation that maps
spins to one-dimentional spinless fermions with creation and
annihilation operation $a_j$ and $a_j^\dagger$ via the relations,
$a_j=(\prod_{l<j} \sigma_l^z)\sigma_j^\dagger$ \cite{Lieb,Sachdev}.
Due to the (quasi) translational symmetry of the system we may
introduce Fourier transforms of the fermionic operator described by
$d_k=\frac{1}{\sqrt{N}}\sum_j a_j \exp(-i2\pi jk/N)$ with
$k=-M,\cdots,M$. The Hamiltonian $H_\phi$ can be diagonalized by
transforming the fermion operators in momentum space and then using
the Bogoliubov transformation. The result is given by $H=\sum_k
\Lambda_k (c_k^\dagger c_k-1)$, where the energy spectrum
$\Lambda_k=\sqrt{(\lambda-\cos(2\pi k/N))+\gamma^2\sin^2(2\pi k/N)}$
and $c_k=d_k \cos\frac{\theta_k}{2}-id_{-k}^\dagger e^{2i\phi}
\sin\frac{\theta_k}{2}$ with the angle $\theta_k$ defined by
$\cos\theta_k=(\cos\frac{2\pi k}{N}-\lambda)/\Lambda_k$.

The ground state $|g\rangle$ of $H_\phi$ is the vacuum of the
fermionic modes described by $c_k |g\rangle=0$, and can be written
as $ |g\rangle=\prod_{k=1}^M \left (
\cos\frac{\theta_k}{2}|0\rangle_k |0\rangle_{-k} -i e^{2i\phi}
\sin\frac{\theta_k}{2}|1\rangle_k|1\rangle_{-k} \right ),$ where
$|0\rangle_k$ and $|1\rangle_k$ are the vacuum and single excitation
of the $k$th mode, $d_k$, respectively. The ground state is a tensor
product of states, each lying in the two-dimensional Hilbert space
spanned by $|0\rangle_k|0\rangle_{-k}$ and
$|1\rangle_k|1\rangle_{-k}$. The geometric phase of the ground
state, accumulated by varying the angle $\phi$ from $0$ to $\pi$, is
described by $\beta_g=-\frac{i}{M}\int_0^\pi \langle g|\partial_\phi
|g\rangle d\phi$\cite{Carollo},  and is found to be

\begin{equation}
\label{Phase} \beta_g=\frac{\pi}{M}\sum_{k=1}^M\left
(1-\frac{\cos(2\pi k/N)-\lambda}{\Lambda_k}\right).
\end{equation}

To study the quantum criticality, we are also interested in the
thermodynamic limit when the spin lattice number $N\ \to \infty$. In
this case the summation $\frac{1}{M}\sum_{k=1}^M$ can be replaced by
the integral $\frac{1}{\pi}\int_0^\pi d\varphi$ with
$\varphi=\frac{2\pi k}{N}$; the geometric phase in the thermodynamic
limit is given by
\begin{equation}
\label{Limit} \beta_g=\int_0^\pi \left
(1-\frac{\cos\varphi-\lambda}{\sqrt{(\cos\varphi-\lambda)^2+\gamma^2\sin^2\varphi}}\right)d\varphi.
\end{equation}

To demonstrate the relation between geometric phase and quantum
phase transitions, we plot geometric phase $\beta_g$ and its
derivative $d\beta_g/d\lambda$ with respect to the field strength
$\lambda$ as a function of the Hamiltonian parameters $\lambda$ and
$\gamma$ in Fig. 1. Two particular features are notable: (i) The
non-analytic property of the geometric phase along the whole
critical line $\lambda_c=1$ in the XY model is clearly shown by
anomalies for the derivative of geometric phase along the same line;
(ii) Geometric phase for the XX model under the thermodynamic limit
is very special in the sense that, instead of using the geometric
phase difference between the ground state and the excited phase as
the signature of phase transition\cite{Carollo}, the geometric phase
of the ground state itself also serves the same role. Geometric
phase under the thermodynamic limit can be obtained explicitly from
Eq.(\ref{Limit}) for $\gamma=0$ as $\beta_g=2\pi-2\arccos(\lambda)$
when $\lambda \leq 1$ and $\beta_g=2\pi$ when $\lambda > 1$. Thus, a
nontrivial geometric phase of the ground state  itself is also a
witness of quantum phase transition; this fact has also been shown
in Ref.\cite{Hamma}.
\begin{figure}[tbph]
\centering
\includegraphics[height=5cm,width=8cm]{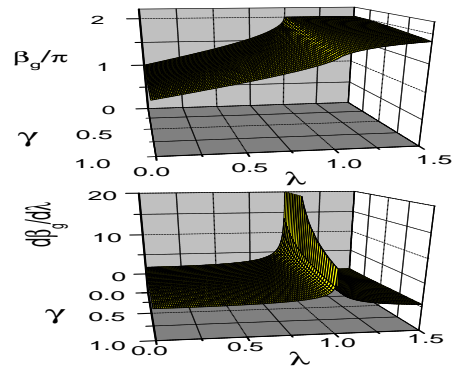}
\caption{(color online). (a) Geometric phase $\beta_g$ of the ground
state (b) and its derivative $d\beta_g/d\lambda$ as a function of
the Hamiltonian parameters $\lambda$ and $\gamma$. The lattice size
$N=10001$. There are clear anomalies for the derivative of geometric
phase along the critical line $\lambda_c=1$. } \label{Fig1}
\end{figure}

\begin{figure}[tbph]
\centering \vspace{1.5cm}
\includegraphics[height=3.5cm,width=8cm]{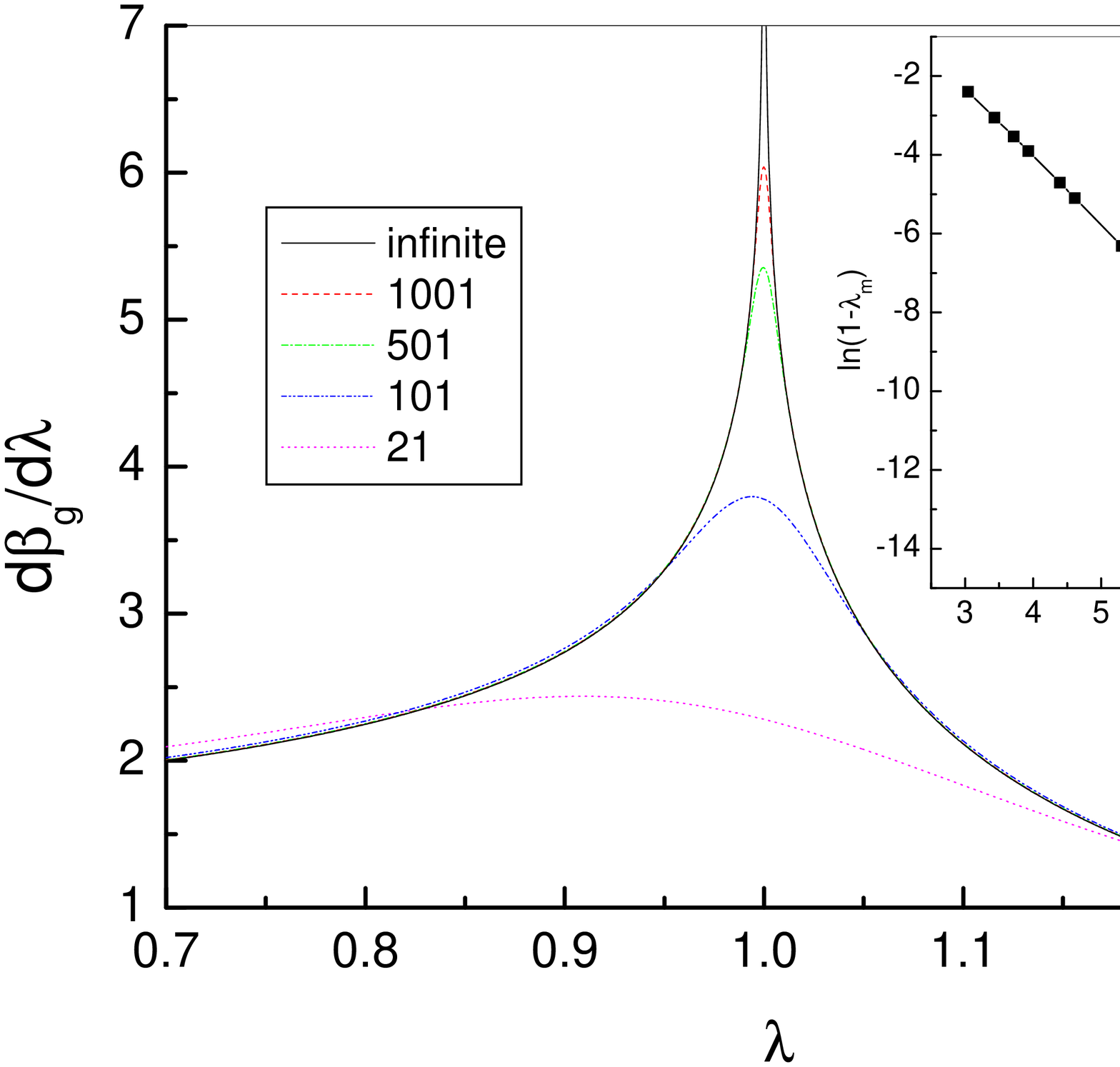}
\vspace{-1.0cm}
\caption{(color online). The derivatives $d\beta_g/d\lambda$ for the
Ising model ($\gamma=1$) as a function of the Hamiltonian parameter
$\lambda$. The curves correspond to different lattice sizes
$N=21,101,501,1001,\infty$. With increasing the system sizes, the
maximum becomes more pronounced. The inset shows that the position
of the maximum changes and tends as $N^{-1.803}$ towards the
critical point $\lambda_c=1$. } \label{Fig2}
\end{figure}

To further understand the relation between geometric phase and
quantum criticality, we investigate the scaling behavior of
geometric phases by the finite size scaling approach\cite{Barber}.
We first look at the Ising model. The derivatives
$d\beta_g/d\lambda$ for $\gamma=1$ and different lattice sizes are
plotted in Fig.\ref{Fig2}. There is no real divergence for finite
$N$, but the curves exhibit marked anomalies and the height of which
increases with lattice size. The position $\lambda_m$ of the peak
can be regarded as a pseudo-critical point \cite{Barber} which
changes and tends as $N^{-1.803}$ towards the critical point and
clearly approaches $\lambda_c$ as $N \to \infty$. As shown in Fig.
3(a), the value of $d\beta_g/d\lambda$ at the point $\lambda_m$
diverges logarithmically with increasing lattice size as:
\begin{equation}
\label{Scaling1} \frac{d\beta_g}{d\lambda}|_{\lambda_m} \approx
\kappa_1 \ln N +\mbox{const.},
\end{equation}
with $\kappa_1=0.3121$. On the other hand, as shown in Fig. 3(b),
the singular behavior of $d\beta_g/d\lambda$ for the infinite Ising
chain can be analyzed in the vicinity of the quantum criticality,
and we find the asymptotic behavior as
\begin{equation} \label{Scaling2}
\frac{d\beta_g}{d\lambda}\approx \kappa_2
 \ln |\lambda-\lambda_c|+\mbox{ const.}, \end{equation}
with $\kappa_2=-0.3123$. According to the scaling ansatz in the case
of logarithmic divergence \cite{Barber}, the ratio
$|\kappa_2/\kappa_1|$ gives the exponent $\nu$ that governs the
divergence of the correlation length. Therefore, $\nu \sim 1$ is
obtained in our numerical calculation for the Ising chain, in
agreement with the well-known solution of the Ising model
\cite{Lieb}. Furthermore, by proper scaling and taking into account
the distance of the maximum of $\beta_g$ from the critical point, it
is possible to make all the data for the value of
$F=[1-\exp(d\beta_g/d\lambda-d\beta_g/d\lambda|_{\lambda_m})]$ as a
function of $N^{1/\nu}(\lambda-\lambda_c)$ for different $N$
collapse onto a single curve \cite{Barber,Osterloh}. The result for
several typical lattice sizes in the Ising model is shown in Fig. 4,
where we can also extract the critical exponent $\nu=1$.

\begin{figure}[tbph]
\centering
\includegraphics[height=5cm,width=8cm]{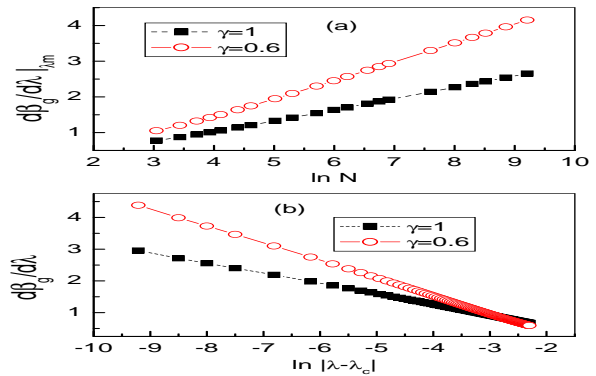}
\caption{(color online). (a) The maximum value of the derivative
$d\beta_g/d\lambda$ at the pseudo-critical point $\lambda_m$  as a
function of lattice sizes. The slope of the line is 0.3121 (0.5234)
for $\gamma=1$ ($\gamma=0.6$). (b) The derivatives
$d\beta_g/d\lambda$ for the thermodynamic limit  logarithmically
diverge on approaching the critical value. The slope of the line is
-0.3123 (-0.5238) for $\gamma=1$ ($\gamma=0.6$).   The ratio between
the two slopes in (b) and (a) for a fix parameter $\gamma$ is the
critical exponent $\nu$. Here $\nu \sim 1$ is obtained for both
$\gamma=0.6$ and $1$, as expected by the concept of universality in
the XY model. } \label{Fig3}
\end{figure}

\begin{figure}[tbph]
\centering
\includegraphics[height=3.5cm,width=8cm]{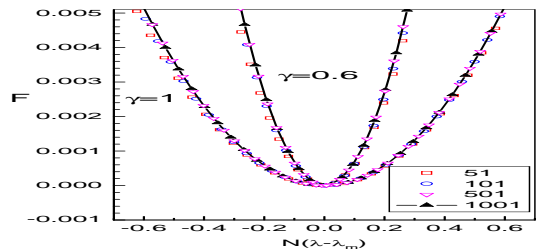}
\caption{(color online). The value of
$F=[1-\exp(d\beta_g/d\lambda-d\beta_g/d\lambda|_{\lambda_m})]$ as a
function $N(\lambda-\lambda_c)$ for different lattice sizes
$N=51,101,501,1001$. All the data for a fix parameter $\gamma$
collapse on a single curve, as expected from the finite size scaling
ansatz. } \label{Fig4}
\end{figure}

A cornerstone of quantum phase transitions is a universality
principle in which the critical behavior depends only on the
dimension of the system and the symmetry of the order parameter. The
XY model for the interval $\gamma \in (0,1]$ belong to the same
universality class with critical exponent $\nu=1$. To verify the
universality principle in this model, we check the scaling behavior
for different values of the parameter $\gamma$. The asymptotic
behaviors are also described by Eqs. (\ref{Scaling1}) and
(\ref{Scaling2}). For instance, from Fig. 3 we get $\kappa_1 \sim
0.5234$ and $\kappa_2=-0.5238$ for $\gamma=0.6$. Moreover, we also
verify that, by proper scaling, all data for different $N$ but a
specific $\gamma$ can collapse onto a single curve. The data for
$\gamma=0.6$ are show in Fig. 4. We can extract the same critical
exponent $\nu=1$ from the data shown in both Fig. \ref{Fig3} and
\ref{Fig4}.

Comparing with the $\gamma\not= 0$ case, the nature of the
divergence of $d\beta_g/d\lambda$ at $\gamma=0$ belongs to a
different universality class, and the scaling behavior of geometric
phase can be directly extracted from the explicit expression of
geometric phase $\beta_g=2\pi-2\arccos(\lambda)$ $(\lambda \leq 1)$
in the thermodynamic limit. Since
$d\beta_g/d\lambda=\sqrt{2}(1-\lambda)^{-1/2}$ $(\lambda \to
1^{-})$, we can infer the known result that the critical exponent
$\nu=1/2$ for the XX model. Therefore, the above results clearly
show that all the key ingredients of the quantum criticality are
present in the geometric phases of the ground state in the XY model.

We now demonstrate that the relation between geometric phase and
quantum phase transitions addressed above is valid in a general
case: quantum phase transition occurs at level crossings or avoided
level crossings, and these kinds of level structures usually can be
captured by the geometric phase of the ground state.
 Consider a generic system described by the Hamiltonian $H(\eta)$ with
$\eta$ a dimensionless coupling constant. For any reasonable $\eta$,
all observable properties of the ground state of $H$ will vary
smoothly as $\eta$ is varied. However, there may be special points
denoted as $\eta_c$, where there is a non-analyticity in some
property of the ground state at zero temperature, $\eta_c$ is
identified as the position of a quantum phase transition.
Non-analytical behavior generally occur at level crossings or
avoided level crossings\cite{Sachdev}. On the other hand, we also
consider geometric phases in a generic many-body system where the
Hamiltonian  can be changed by varying the parameters ${\bf R}$ on
which it depends. The state $|\psi (t)\rangle$ of the system evolves
according to Schrodinger equation $H ({\bf R}
(t))|\psi(t)\rangle=i\hbar
\partial_t |\psi (t)\rangle$. At any instant, the natural basis
consists of the eigenstates $|n({\bf R})\rangle$ of $H({\bf R})$ for
${\bf R}={\bf R}(t)$, that satisfy $H ({\bf R})|n ({\bf
R})\rangle=E_n ({\bf R})|n ({\bf R})\rangle$ with energy $E_n ({\bf
R})$ $(n=1,2,3\cdots)$. Berry showed that the geometric phase for a
specific eigenstate, such as the ground state
($|g\rangle=|1\rangle$) of a many-body system we concern here,
adiabatically undergoing a closed path in parameter space denoted by
$C$, is given by $\beta_g (C)=-\int\int_C V_g ({\bf R})\cdot d{\bf
S}$, where $d{\bf S}$ denotes area element in ${\bf R}$ space and
$V_g ({\bf R})$ is the Berry curvature given by\cite{Berry}
\begin{equation}
\label{Curvarure} V_g ({\bf R})=Im\sum_{n \not= g}\frac{\langle
g|\nabla_{\bf R} H|n\rangle \langle n|\nabla_{\bf R}
H|g\rangle}{(E_n-E_g)^2}.
\end{equation}
The energy denominators in Eq.(\ref{Curvarure}) show that the Berry
curvature usually diverges at the point of parameter space where
energy levels are cross and may have maximum values at avoided level
crossings. Thus level crossings or avoided level crossings, the two
specific level structures related to quantum phase transitions, are
reflected in the geometry of the Hilbert space of the system and can
be captured by geometric phases of the ground state. Therefore, the
relation between geometric phase and quantum phase transitions
demonstrated herein is in fact a very general result and not a
specific property of the XY model.

In summary, we established the close relation between geometric
phase of the ground state and quantum phase transitions in a general
many-body system. As a typical example,  we show in detail that all
the key ingredients of quantum criticality, such as scaling
features, critical exponents and universality, etc., are present in
the geometric phases in the XY spin chain.

I thank L. M. Duan, H. Fu, J. K. Pachos, and P. Zanardi for helpful
discussions and P. Berman for his critical reading of this paper.
This work was supported by NSF FOCUS center, the MCTP, the NCET and
NSFC under grant No. 10204008.

{\sl Note added}-- After this paper was completed, I got a
manuscript\cite{Hamma} where the general relation among Berry
phases, topology, and quantum phase transitions in many body systems
was independently established.

\end{document}